\documentstyle[psfig]{mn}

\title{High--resolution simulations of 
stellar collisions between equal-mass main-sequence stars in globular clusters}
\author[Alison Sills, Tim Adams, Melvyn Davies, Matthew Bate ]
{Alison Sills \thanks{Current Address: Department of Physics and
Astronomy, McMaster University, 1280 Main St. W., Hamilton, ON, L8S
4M1, Canada}, Tim Adams, Melvyn B. Davies \\ Department of Physics and
Astronomy, University of Leicester, Leicester, LE1 7RH
\and
 Matthew R. Bate \\ School of Physics, University of Exeter, Exeter, EX4 4QL}

\begin{document}
\maketitle

\begin{abstract}

We performed high--resolution simulations of two stellar collisions
relevant for stars in globular clusters. We considered one head-on
collision and one off-axis collision between two 0.6 $M_{\odot}$ main
sequence stars.  We show that a resolution of about 100 000 particles
is sufficient for most studies of the structure and evolution of blue
stragglers. We demonstrate conclusively that collision products
between main-sequence stars in globular clusters do not have surface
convection zones larger than 0.004 $M_{\odot}$ after the collision,
nor do they develop convection zones during the `pre-main-sequence'
thermal relaxation phase of their post-collision evolution. Therefore,
any mechanism which requires a surface convection zone (i.e. chemical
mixing or angular momentum loss via a magnetic wind) cannot operate in
these stars. We show that no disk of material surrounding the
collision product is produced in off-axis collisions. The lack of both
a convection zone and a disk proves a continuing problem for the
angular momentum evolution of blue stragglers in globular clusters.

\end{abstract}


\section{Introduction and Motivation}\label{intro}

It has been shown that the products of main-sequence -- main-sequence
collisions appear in the colour-magnitude diagrams of clusters as blue
stragglers \cite{SLBDRS97,SBH97}. Since blue stragglers are readily
observable in clusters, they form an ideal population with which to
probe the dynamical evolution of the cluster. The dynamical state of a
globular cluster (its density profile, velocity dispersion, amount of
mass segregation etc.) will determine the rate and nature of the
collisions which occur in the cluster. As the cluster evolves, the
kinds of collision that occur will change. Therefore, the population
of collision products in a cluster can be used to probe the history of
the cluster \cite{SBEG00,HTAP01}.  However, in order to use collision
products in this way, there are two issues which must first be
understood. First, we know that blue stragglers can also be formed
through the merger of two components of a binary system. These blue
stragglers will probably have different properties than those formed
from collisions. We must either be confident that the population we
are observing is collisional in origin (e.g. from cluster density
considerations), or be able to distinguish between the two
populations. Secondly, we also need to be sure that we understand the
formation and evolution of the collision products themselves. This
paper is concerned with addressing the second point.

In this paper, we present the highest resolution smoothed particle
hydrodynamic (SPH) simulations of collisions between main-sequence
stars to date. Most recent computations have $\sim 10^4$ particles
\cite{LRS96,SBH97}, with the highest resolution simulation using $10^5$
particles \cite{SFLRW01}. In this paper, we increase the number of
particles to $10^6$. There are three main reasons to extend this kind
of simulation to such high resolution. The first, and simplest, is to
make sure that no fundamental change in our understanding of blue
stragglers occurs. In earlier work, Benz \& Hills \shortcite{BH87}
performed an SPH simulation using 1024 particles, and concluded that
collision products are fully mixed (i.e. the resulting star is
chemically homogeneous). Ten years later, these simulations were
repeated but with a factor of 10-50 more particles
\cite{LRS96}. Because of the higher resolution, it became clear that
collision products are NOT chemically homogeneous, but instead retain
some memory of the chemical profile of the parent stars of the
collision. We feel confident that such a fundamental change in our
understanding will not happen again when we increase the resolution,
but we want to be sure.

More importantly, we performed these high resolution calculations to
answer two fundamental questions about the structure of the collision
products immediately after the collision. The first involves the
structure of the outer layers of the star, and the second is concerned
with the material which is thrown off by the star during the collision
itself. There has been some debate in the literature recently about
whether the collision product has a surface convection zone shortly
after the end of the collision. The presence of such a convection zone
could have ramifications for both the surface chemical abundances and
the rotation rate of the star when it reaches its main sequence
\cite{LL95}. Previous SPH simulations and subsequent evolution
calculations have shown that no convection zone exists at the end of
the collision, nor does it appear during the initial thermal
relaxation of the star \cite{LRS96,SLBDRS97}.  However, members of the
community argue that the previous simulations are unable to resolve
the outermost regions of the star well at all, due to their low
particle number and use of equal-mass particles (Livio, private
communication).  In this paper, we resolve this issue by increasing
the resolution of our simulations in the outer regions of the parent
stars in particular, and by using variable mass particles.

We are also interested in following the material which is thrown off
from the stars during the collision. We have discovered, using
previous generation simulations, that blue stragglers which are formed
by an off-axis collision have an angular momentum problem
\cite{SFLRW01}. Namely, these stars retain too much of their angular
momentum, and have no apparent way of losing it during their thermal
relaxation phase after the collision. As a result, the collision
products spin up during their collapse to the main sequence, and
inevitably rotate faster than their break-up velocity. Such collision
products can never become blue stragglers, since they tear themselves
apart before they reach the main sequence. We wish to follow the lost
material with greater resolution, to study the amount of angular
momentum that this material carries away with it. We also wish to
follow the outermost material of the parent stars to see if some of it
forms a disk around the collision product. If so, then we can
plausibly suggest that magnetic locking to a disk is responsible for
removing angular momentum from the collision product.

In \S\ref{method} we describe the methods used to model the stellar
collisions and their subsequent evolution. We present our results in
\S\ref{results}, and discuss their implications in \S\ref{summary}.

\section{Method}\label{method}

The simulations of stellar collisions discussed in this paper were
performed using the smoothed particle hydrodynamics (SPH) method
\cite{B90,M92}. Our three dimensional code uses a tree to solve for the
gravitational forces and to find the nearest neighbours
\cite{BBCP90}. We use the standard form of artificial viscosity with
$\alpha=1$ and $\beta=2.5$, and an adiabatic equation of state. The
thermodynamic quantities are evolved by following the change in
internal energy. Both the smoothing length and the number of
neighbours can change in time and space. The smoothing length is
varied to keep the number of neighbours approximately constant ($\sim$
50 for the low resolution runs, and $\sim$ 200 for the million
particle run). The code is the parallel version of the code described
in detail in Bate, Bonnell \& Price \shortcite{BBP95}. It was
parallelized using OpenMP, and was run on the SGI Origin 3000 operated
by the UK Astrophysical Fluids Facility (UKAFF) based at the
University of Leicester.

We modelled collisions between two equal--mass stars. We chose a mass
of 0.6 $M_{\odot}$ as a representative, but not extreme, mass for
globular cluster stars. The initial stellar models were calculated
using the Yale stellar evolution code (YREC, Guenther et
al. 1992), and had a metallicity of Z=0.001 and an age
of 15 Gyr. If the goal of the simulation is to investigate the
detailed structure of the collision product, then it is crucial to
begin with a realistic stellar model rather than a polytrope or other
approximation \cite{SL97}. The particles were initially distributed on
an equally--spaced grid, and their masses were varied until the
density profile matched that of the stellar model. By using unequal
mass particles, we increase the resolution in the outer, low density
regions of the star. This is important for resolving the outer layers
of the collision products (to determine if there is a convection zone
or not) and for following the material which is thrown off during the
collision.

The stars were given a relative velocity at infinity of
$v_{\infty}$=10 km/s, which is a reasonable value for globular cluster
stars. They were set up on almost parabolic orbits with a pericentre
separation $r_p=0$ for the head-on collisions, and $r_p$=0.25
$R_{\odot}$ for the off-axis collisions. (The radius of the main
sequence star is $0.61 R_{\odot}$.) Most collisions are expected to
happen during interactions involving binary stars \cite{HB83}, so the
stars are on bound, and usually highly elliptical, orbits.  For both
the head-on and off-axis collisions, the number of particles was
varied between 10 000 and 300 000. We also ran one head-on collision
with $10^6$ particles. The details of the collisions are shown in
Table \ref{description}. We give a single letter name for each run,
the pericentre separation $r_p$ in solar radii, the number of
particles, the initial value of the smoothing length $h_0$ in solar
radii, the minimum and <maximum particle masses in solar masses, and
the amount of mass lost from the system during the collision in solar
masses.

\begin{table*}
\begin{minipage}{200mm}
\caption {Description of SPH simulations performed \label{description}}
\begin{tabular}{|ccrlccc|}
\hline
Run &  $r_p$ & $N_{\rm part}$ & $h_0$ & $M_{\rm min}$ & $M_{\rm max}$ & $M_{\rm lost}$  \\
    &  $R_{\odot}$ & & $R_{\odot}$ & $M_{\odot}$ & $M_{\odot}$ & $M_{\odot}$\\
\hline
A & 0.0  &  10 162 & 0.058 & $2.10 \times 10^{-6}$ & $4.58 \times 10^{-3}$ & $8.14 \times 10^{-2}$ \\
B & 0.0  &  29 950 & 0.0417 & $3.71 \times 10^{-7}$ & $1.37 \times 10^{-3}$ & $7.76 \times 10^{-2}$  \\
C & 0.0  & 100 034 & 0.028568 & $4.96 \times 10^{-8}$ & $4.47 \times 10^{-4}$ & $7.49 \times 10^{-2}$  \\
D & 0.0  & 299 398 & 0.02011 & $5.27 \times 10^{-9}$ & $1.56 \times 10^{-4}$ & $7.41 \times 10^{-2}$  \\
E & 0.0  & 999 778 & 0.0136 & $2.98 \times 10^{-10}$ & $4.75 \times 10^{-5}$ & $7.47 \times 10^{-2}$ \\
F & 0.25 &  10 162 & 0.058 & $2.10 \times 10^{-6}$ & $4.58 \times 10^{-3}$ & $3.45 \times 10^{-2}$ \\ 
G & 0.25 &  29 950 & 0.0417 & $3.71 \times 10^{-7}$ & $1.37 \times 10^{-3}$ &  $3.62 \times 10^{-2}$  \\
H & 0.25 & 100 034 & 0.028568 & $4.96 \times 10^{-8}$ & $4.47 \times 10^{-4}$ & $3.99 \times 10^{-2}$  \\
I & 0.25 & 299 398 & 0.02011  & $5.27 \times 10^{-9}$ & $1.56 \times 10^{-4}$ & $4.16 \times 10^{-2}$  \\
\hline
\end{tabular}
\end{minipage}
\end{table*}

The results of these three dimensional SPH simulations were converted
to one dimensional models suitable for starting models for the Yale
stellar evolution code, YREC. The entropy of the particles and their
chemical abundances were averaged over surfaces of constant
gravitational potential and binned into $\sim$ 100 bins. The structure
of the collision product is determined from these profiles using the
equation of hydrostatic equilibrium. The temperature profile is
calculated using the ideal gas equation of state.  For more details of
the conversion between SPH results and stellar evolution models, see
Sills et al. \shortcite{SLBDRS97,SFLRW01}.

We used YREC to calculate the evolution of the collision
products. Using the starting models described above, we followed their
evolution during the thermal relaxation phase (analogous to the
pre-main-sequence of normal stars), through their main-sequence
lifetimes and to the giant branch.

\section{Results}\label{results}

\begin{figure}
\psfig{file=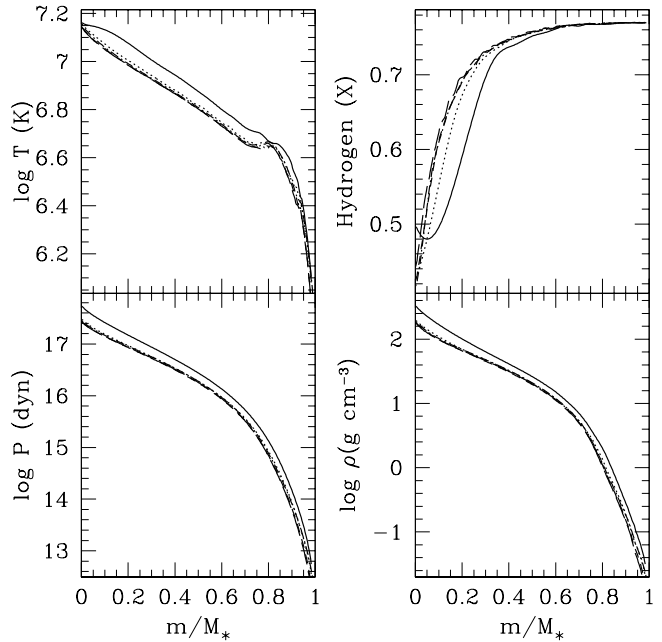,width=9cm}
\caption {Structure of head-on collision products. The main structure
parameters of a non-rotating star have been plotted as a function of
mass fraction: pressure, temperature, density, and hydrogen mass
fraction. The different line styles show the results for different
numbers of SPH particles. In increasing order of particles, they are:
solid (10 000 particles), dotted (30 000 particles), short-dashed (100
000 particles) long-dashed (300 000 particles), and dot-dashed (1 000
000 particles). The models converge towards the 1 000 000 particle
simulation, and there is very little obvious difference between the
100 000, 300 000 and 1 000 000 particle results. \label{structure}}
\end{figure}

We performed five simulations of head-on stellar collisions, with the
number of particles ranging from $10^4$ to $10^6$. The average
resolution elements for these collisions ranged from $2h=0.116
R_{\odot}$ to $2h=0.027 R_{\odot}$ respectively.  Figure
\ref{structure} shows the profiles of the important structural
parameters as a function of mass fraction in the product of a head-on
collision between two 0.6 M$_{\odot}$ stars. The different line styles
denote increasing particle numbers in the following order: 10 000
(solid), 30 000 (dotted), 100 000 (short dashed), 300 000 (long
dashed), 1 000 000 (dot--dashed). The low resolution simulations do a
reasonable job of producing the correct structure for the collision
products. However, as the resolution increases, the models converge to
an answer different from that seen in the low resolution runs.  This is
best seen in the plot of hydrogen abundance as a function of mass
fraction. The 100 000 and 300 000 particle simulations have converged
to a solution for the structure of this collision product.  The higher
resolution simulations also produce smoother profiles. This is
expected, since we have the information from more particles
contributing to the values of the structural quantities (pressure,
density, etc.) in each shell. Therefore, we suggest that the optimum
number of particles for studying the structure of stellar collision
products is around 100 000.

\begin{figure}
\psfig{file=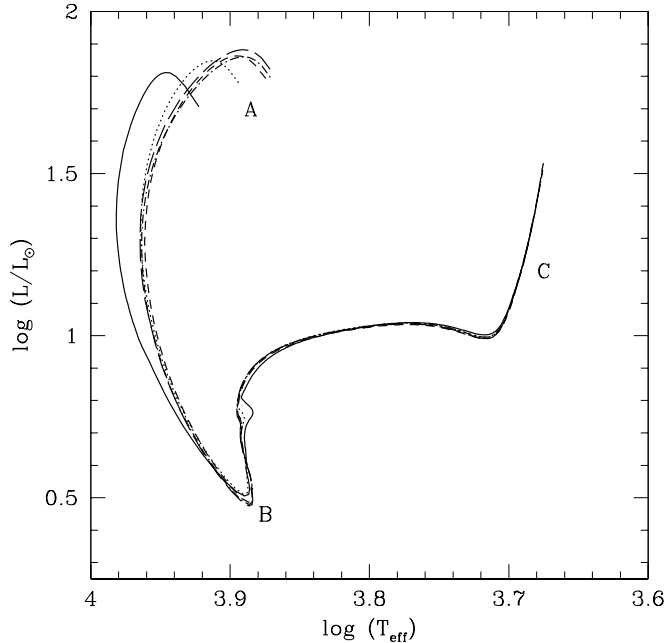,width=9cm}

\caption {Evolutionary tracks for head-on collision products. The stars 
begin their post-collision evolution far from the Hayashi track, but
are still bright (position A). As they contract, they move towards the
main sequence (position B), and then follow a standard evolutionary
track towards the giant branch (position C).  The line styles are the
same as in Figure \ref{structure}. The difference in structure between
the different resolution simulations manifests itself as a difference
in evolutionary track shape. The tracks are most different during
their initial contraction, and then again near the turnoff of the main
sequence. The lowest resolution simulation (10 000 particles, solid
line) has a hook in its evolutionary track near the turnoff, evidence
of a well-developed core convection zone. The higher resolution
simulations do not show this feature. \label{tracks}}
\end{figure}

In Figure \ref{tracks}, we present the evolutionary tracks which
result from the starting models shown in Figure \ref{structure}. The
stars begin their post-collision lives at position A. They are quite
bright, and are significantly out of thermal equilibrium because of
the energy deposited into their envelopes by the collision. They
contract to the main sequence (position B), where core hydrogen
burning begins. The stars look like normal main-sequence stars, except
that their central hydrogen abundance is less than it would be for a
normal star of that mass. Therefore, their main-sequence lifetime is
shorter than that of a normal 1.1 $M_{\odot}$ star (3.3 Gyr). The
collision products reach the turnoff, where hydrogen is exhausted in
their core, and evolve towards the giant branch (position C), exactly
like normal stars. The time since the collision at three benchmark
positions along the evolutionary track is given in Table
\ref{timescales}.  Both the post-collision structure and the
evolutionary tracks agree with previous simulations of the same
collision \cite{SLBDRS97}.

\begin{table}
\caption {Evolutionary Timescales (Gyr) \label{timescales}}
\begin{tabular}{|ccccc|}
\hline
Run &  Return to MS & Turnoff & Giant Branch \\
\hline
A & 1.60 $\times 10^{-3}$ & 1.41 & 2.08 \\
B & 1.90 $\times 10^{-3}$ & 1.64 & 2.41 \\
C & 2.20 $\times 10^{-3}$ & 1.85 & 2.69 \\
D & 2.16 $\times 10^{-3}$ & 1.92 & 2.79 \\
E & 2.20 $\times 10^{-3}$ & 1.87 & 2.73 \\
\hline
\end{tabular}
\end{table}

The different line styles in Figure \ref{tracks} indicate the
different number of particles in the simulation, with the line styles
having the same meaning as in Figure \ref{structure}. We see the same
trend with increasing resolution of the simulation -- the tracks
converge to a common solution around 100 000 particles. This is
expected, since the evolution of a star is determined by its
structure. The same trend of convergence is seen in the evolutionary
timescales in Table \ref{timescales}. Note that the largest differences
between simulations show up in the thermal relaxation phase, and that
by the main sequence, all the simulations are almost
indistinguishable. The exception is the lowest resolution
simulation. The solid black line begins its main sequence quite close
to the dotted line (30 000 particle simulation), but has a hook
near the turnoff, indicating the presence of a central convection
zone. As the resolution increases, we see that this convection zone
disappears. Simulations with $10^4$ particles are not quite detailed
enough to accurately depict the true structure of these collision
products.

\begin{figure}
\psfig{file=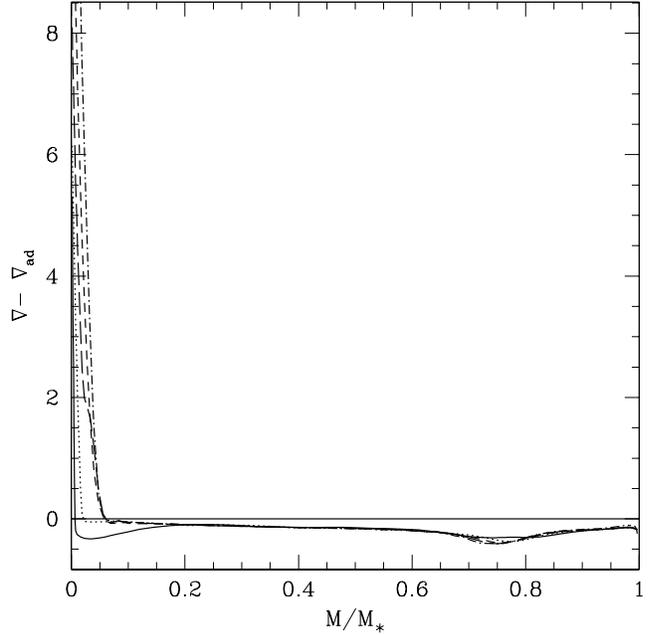,width=9cm}

\caption{Convective stability criterion $\nabla-\nabla_{ad}$ as a function
of mass fraction for the head-on collision models immediately after
the collision. The line styles are the same as in Figure
\ref{structure}. When $\nabla-\nabla_{ad}$ is less than zero, the region
is stable against convection. Except for a small region at the centre
of the star, the entire collision product is stable against
convection. \label{convec}}
\end{figure}

The criterion for convective stability in stellar models is the
Schwarzschild criterion:

\begin{equation}
\nabla - \nabla_{ad} \leq 0
\end{equation}
where $\nabla_{ad}$ is the adiabatic temperature gradient in the star,
and $\nabla$ is the radiative temperature gradient:
\begin{equation}
\nabla = \frac{d \ln T}{d \ln P} = \frac{3}{64 \pi \sigma G} \frac{\kappa L P}{M T^4}
\end{equation}

where $\sigma$ is the Stefan Boltzmann constant, $G$ is the
gravitational constant, and $\kappa, L, P, M$ and $T$ are the opacity,
luminosity, pressure, enclosed mass and temperature at that position
in the star. The models taken from the SPH results were used as
starting models in the stellar evolution code, and allowed to relax,
so that all the equations of stellar structure are satisfied. Since
our SPH code does not allow for energy transport, the luminosity
distribution in the star can only be determined from the other
structure parameters using the equations of stellar structure. The
temperature gradients were calculated, and their difference is plotted
as a function of mass fraction in Figure \ref{convec}. The line styles
are the same as in Figure \ref{structure}. We can clearly see a
convective core in these stars, where $\nabla - \nabla_{ad} \ > 0$.
However, the entire star outside of the inner 0.1 $M_{\odot}$ is not
convective.  The outermost shell in all our stellar models is at a
mass fraction $M/M_{\star} = 0.99667$. This value is the same for all
models, regardless of the number of SPH particles, because of the way
we have transformed the SPH information to the stellar evolution
starting model.  The total mass of these stars is 1.12 $M_{\odot}$, so
only the outer 0.004 $M_{\odot}$ is not resolved. Therefore, we can
say with certainty that this stellar collision product does not have a
surface convection zone larger than 0.004 $M_{\odot}$. Our
evolutionary models show that the stellar collision products do not
develop surface convection zones during their thermal contraction to
the main sequence.

\begin{figure}
\psfig{file=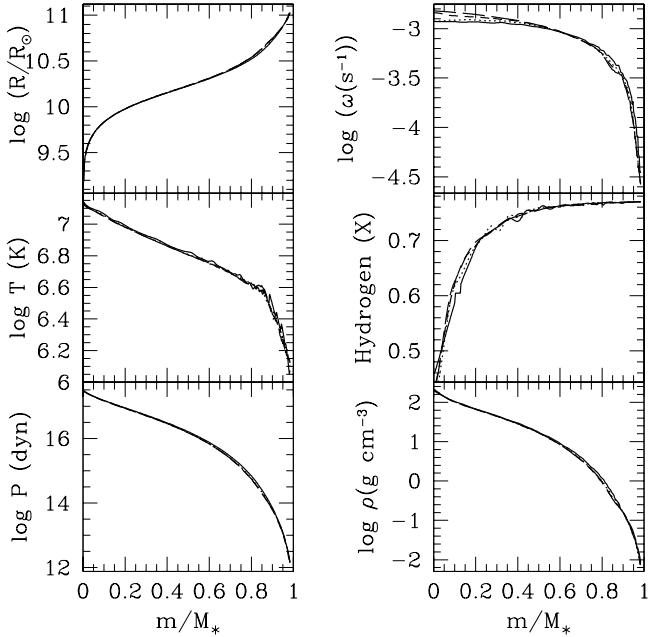,width=9cm}
\caption {Structure of off-axis collision products. The line styles are 
the same as in Figure \ref{structure}. Again, we see a convergence to
the 100 000 particle simulation. \label{structrot}}.
\end{figure}

As well as studying the head-on collisions, we performed one off-axis
collision with a pericentre separation of 0.25 R$_{\odot}$. Figure
\ref{structrot} shows the structure profiles of the collision products
at various resolutions. The line styles are the same as in Figure
\ref{structure}, except that we have not run a million-particle off-axis
simulation. In addition to profiles of pressure, temperature, density
and hydrogen mass fraction, we show the angular velocity and radius
profiles.  Again, there is clearly a convergence of the structural
properties as the number of particles increases. The 100 000 particle
(short dashed line) and 300 000 particle (long dashed line)
simulations are almost indistinguishable.

The final question we wish to investigate with these high resolution
simulations involves the material which is thrown off during the
collision. Between 1 and 10\% of the original stellar material becomes
unbound from the system as it is shock-heated during these kinds of
collisions, depending on the masses of the parent stars and the impact
parameter \cite{LRS96}. Other material is thrown to large distances
but remains bound and falls back onto the collision product. If some
of this material has sufficient angular momentum, it will form a disk
around the collision product. The existence of such a disk could prove
important for transporting angular momentum from the central collision
product.

\begin{figure}
\psfig{file=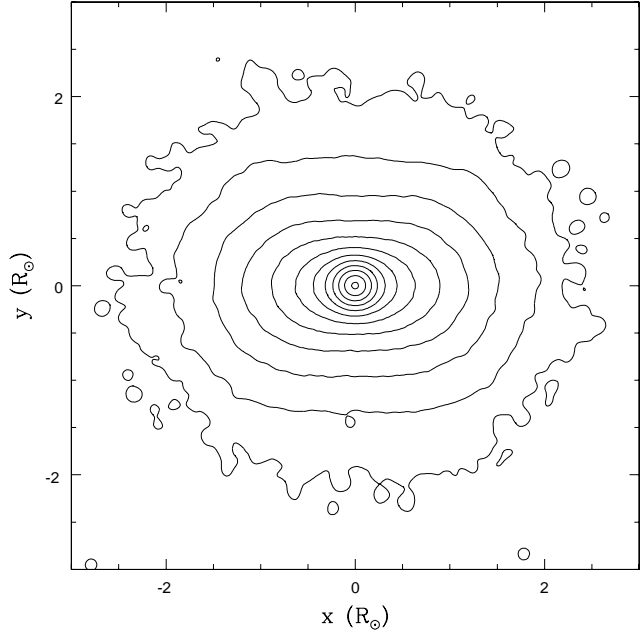,width=9cm}
\caption 
{Density profile of an off-axis stellar collision product between two
0.6 $M_{\odot}$ stars, modelled using 300 000 particles.  This is a
5x5 $R_{\odot}$ box in the x-z plane around the collision product. The
logarithmically spaced density contours are 0.5 dex apart.  The
collision product is rotating around the z axis, and shows significant
flattening due to rotation. However, there is no indication of a
disk. \label{contour}}
\end{figure}

Figure \ref{contour} is a snapshot of the 300 000 particle off-axis
stellar collision taken 50 dynamical times after the collision
began. The star has reached hydrostatic equilibrium and is settled
down to its post-collision state.  The contours show the density of
the material in a slice through the x-z plane of the collision
product. The collisions were constrained to occur in the x-y plane, so
the product is rotating around the z axis. It is rotating rapidly, and
therefore is highly flattened by rotation. However, there is no
indication that a circumstellar disk is present. This is in agreement
with the results of Benz \& Hills (1987), who find that disks are
present only for simulations with larger impact parameters. 

\section{Summary and Discussion}\label{summary}

We have performed the highest resolution SPH simulations of collisions
between main-sequence stars to date, using up to $10^6$ SPH
particles. The collisions were between equal mass main-sequence stars
in globular clusters -- the stars have masses of 0.6 $M_{\odot}$,
metallicity Z=0.001, age of 15 Gyr and a relative velocity at infinity
of 10 km/s. We varied the resolution of our simulations from $10^4$ to
$10^6$ SPH particles, and we looked at a head-on collision and an
off-axis collision. These simulations took between $\sim 6$ CPU
hours and $\sim 5000 $ CPU hours, and the CPU time scaled approximately
linearly with N $\log$ N.  We had three goals for performing this research.

Our first goal was to determine the optimal resolution for SPH
simulations of this kind. Since we want to use the results of the SPH
simulations to study the detailed structure and evolution of the
collision products, we need to have an accurate model of the
distribution of the SPH particles and their properties. On the other
hand, we do not want to waste valuable computer resources going to a
higher resolution than is necessary. We have shown that both head-on
and off-axis simulations with $\sim$ 100 000 particles give
essentially the same information as runs with higher resolution, and
therefore we suggest that most simulations with similar requirements
can confidently use on order $10^5$ particles.

Our second goal was to use these high resolution simulations to study
the outer few percent of the collision products. In particular, we
were interested in the existence of a surface convection zone, for two
reasons. One consequence of a convection zone is mixing of elements in
the convection zone. If the zone reaches deep enough in a star to
dredge up nuclear processed material, for example, the surface
abundances of elements like C, N and O will be different from what we would
expect for primordial material. In blue stragglers and other collision
products, surface abundances could be used to determine, perhaps, the
masses of the parent stars. However, this can only be done if we
understand whether the abundances should be mixed by convection or
not. The second consequence of a surface convection zone is angular
momentum loss. Stars can lose angular momentum through a magnetic wind
\cite{K88}. This process is most effective in stars with deep
convection zones, since they can support stronger magnetic fields and
since the angular momentum is drained out of the entire (uniformly
rotating) convection zone.

We have shown that the product of the head-on stellar collision 
investigated in this paper does not have a surface convection zone 
larger than 0.004 $M_{\odot}$. For comparison, the Sun is considered to 
have a fairly shallow convection zone, with a mass of 0.02 $M_{\odot}$ 
\cite{GDPK92}. A convection zone less than about 0.01 $M_{\odot}$ 
is certainly not deep enough to substantially modify the surface abundances 
of any element, with the possible exception of lithium, beryllium and boron 
(which are destroyed at low temperatures). Angular momentum loss is also 
going to be very small for stars with such small convection zones.

Our final goal was to determine whether the off-axis stellar collision
products had a circumstellar disk. If a rotating star has a disk with
a mass of about 0.01 $M_{\odot}$, and it is locked to the disk by a
magnetic field, then angular momentum will be transported from the
star to the disk by the field \cite{K91}. The net effect of the
angular momentum transport will be to slow the star's rotation
rate. This effect is thought to operate in pre-main-sequence stars
\cite[and references therein]{SPT00}. Unfortunately, our simulations
do not show any evidence for a disk. Based on previous work
\cite{BH87}, we expect that blue stragglers which were formed from
collisions with larger impact parameters will have circumstellar
disks. However, the low impact parameter collision we modelled in this
simulation is still rotating too quickly to contract to the main
sequence without losing some angular momentum.

Sills et al. \shortcite{SFLRW01} showed that blue stragglers that are
formed by stellar collisions have an angular momentum problem. The
collision products are formed with large total angular momentum, and
some of them are rotating near their break-up velocity. The collision
products are large objects, and they start to contract to the main
sequence. As they contract, they spin up since they have no way to
lose any angular momentum. They begin to reach break-up velocities and
become unstable.  Even by shedding material, they cannot remove enough
angular momentum to remain bound, and they tear themselves apart. We
are forced to one of two conclusions; either blue stragglers are not
made by stellar collisions (since head-on collisions are very rare,
and all off-axis collisions have too much angular momentum), or
collision products have some way of losing their angular momentum. The
two standard ways stars are thought to lose angular momentum is
through a magnetic wind, or through disk-locking.  We have shown that
both these scenarios are not viable, since these collision products do
not have surface convection zones or circumstellar disks immediately
after the collision. However, it is interesting to note the work of
Durisen et al. \shortcite{DGTB86}, who show that if a polytrope is
rotating very rapidly, some outer material is thrown off into a disk,
leaving a stable (but non--axisymmetric) central object. In order for
this to happen, the star has to have a ratio of rotational kinetic
energy to gravitational energy $\beta \geq 0.3$. At the end of the
collision, our off-axis collision product has $\beta \sim 0.045$, and
is therefore stable against any bar instability. However, as the star
evolves and contracts, it is possible that $\beta$ could becomes high
enough to trigger some of these hydrodynamic rotational instabilities.

In this paper, we have limited ourselves to two very specific
collisions.  The two stars involved in the collision were of equal
mass, and we only investigated two choices of impact parameter. There
have been other studies of main-sequence star stellar collisions which
covered more parameter space \cite{BH87,LRS96,SBH97}.  These studies
have clearly and carefully outlined the properties of collisions under
different circumstances. These high resolution simulations were meant
to answer some specific questions about the details of stellar
collisions and the resulting collision products. We can generalize our
current results to all main-sequence star collisions in globular
clusters under the limitations outlined in the previous
papers. Therefore, we feel justified in claiming that no collision
products have surface convection zones, and that collisions do not
create long--lived circumstellar disks.  The search to understand the
angular momentum evolution of blue stragglers will have to turn to
other avenues, possibly involving a more detailed look at the
combination of stellar evolution and hydrodynamic instabilities after
the collision.

\section{Acknowledgments}
This work was supported by PPARC. The computations reported here were
performed using the UK Astrophysical Fluids Facility
(UKAFF). Development work was performed using the University of
Leicester Mathematical Modelling Centre's supercomputer which was
purchased through the EPSRC strategic equipment initiative.  MBD
gratefully acknowledges the support of a URF from the Royal Society,
and TA gratefully acknowledges support through a PPARC research
studentship.  The authors would like to thank the referee, Marc
Freitag, for his helpful reading of the manuscript.


\begin{thebibliography}{}

\bibitem[\protect\citename{Bate et al. } 1995]{BBP95} Bate, M.R., Bonnell, 
  I.A., Price, N.M., 1995, MNRAS, 277,  362

\bibitem[\protect\citename{Benz } 1990]{B90} Benz, W., 1990, in Buchler 
  J. R., ed., The Numerical Modeling of Nonlinear Stellar Pulsations: 
  Problems and Prospects. Kluwer, Dordrecht, p. 269

\bibitem[\protect\citename{Benz et al. }1990]{BBCP90} Benz, W., Bowers R. L.,
  Cameron A. G. W., Press W., 1990, ApJ, 348, 647

\bibitem[\protect\citename{Benz \& Hills }1987]{BH87} Benz, W., Hills, J. G., 
1987, ApJ, 323, 614

\bibitem[\protect\citename{Durisen et al. }1986]{DGTB86} Durisen, R. H., Gingold, R. A., Tohline, J. E., Boss, A. P., 1986, ApJ, 305, 281

\bibitem[\protect\citename{Guenther et al. }1992]{GDPK92} Guenther, D. B., Demarque, P.,  Pinsonneault, M.\ H., \& Kim, Y-.-C., 1992, ApJ, 392, 328

\bibitem[\protect\citename{Hurley et al. }2001]{HTAP01} Hurley, J.~R., Tout, C.~A., Aarseth, S.~J., \& Pols, O.~R.\ 2001, MNRAS, 323, 630

\bibitem[\protect\citename{Hut \& Bahcall }1983]{HB83} Hut, P., Bahcall, J. B., 1983, ApJ, 268, 319

\bibitem[\protect\citename{Kawaler }1988]{K88} Kawaler, S.D., 1988, ApJ, 333, 236

\bibitem[\protect\citename{Konigl }1991]{K91} Konigl, A., 1991, ApJ, 370, L39

\bibitem[\protect\citename{Leonard \& Livio }1995]{LL95} Leonard, P.J.T., 
Livio, M., 1995, ApJ, 447, L121

\bibitem[\protect\citename{Lombardi et al. }1996]{LRS96} Lombardi, J. C., Jr., 
  Rasio, F. A., Shapiro, S. L., 1996, ApJ, 468, 767

\bibitem[\protect\citename{Monaghan }1992]{M92} Monaghan, J.J, 1992, ARA\&A,
30, 543

\bibitem[\protect\citename{Sandquist et al. }1997]{SBH97} Sandquist, E., 
  Bolte, M., \& Hernquist, L., 1997, ApJ, 477, 335

\bibitem[\protect\citename{Sills \& Lombardi }1997]{SL97} Sills, A., Lombardi, 
J. C., Jr., 1997, ApJL, 484, 51

\bibitem[\protect\citename{Sills et al. }1997]{SLBDRS97} Sills, A.,
  Lombardi, J. C., Jr., Bailyn, C. D., Demarque, P. D., Rasio, F. A.,
  Shapiro, S. L., 1997, ApJ, 487, 290

\bibitem[\protect\citename{Sills, Pinsonneault \& Terndrup }2000]{SPT00} 
  Sills, A., Pinsonneault, M. H., Terndrup, D. M., 2000, ApJ, 534, 335

\bibitem[\protect\citename{Sills et al. }2000]{SBEG00}Sills, A., Bailyn,
  C. D., Edmonds, P. T., Gilliland, R. L., 2000, ApJ, 535, 298

\bibitem[\protect\citename{Sills et al. }2001]{SFLRW01}Sills, A., Faber, 
  J. A., Lombardi, J. C., Jr., Rasio, F. A., Warren, A. R., 2001, ApJ, 548, 
  323

\end{thebibliography}
\end{document}